\pdfoutput=1
\documentclass[paper]{ieice}
\usepackage[pdftex]{graphicx,xcolor}% for pdflatex
\usepackage[fleqn]{amsmath}
\usepackage{newtxtext}
\usepackage[varg]{newtxmath}

\usepackage{subcaption}
\usepackage{amsbsy}
\newcommand{\AmSLaTeX}{%
 $\mathcal A$\lower.4ex\hbox{$\!\mathcal M\!$}$\mathcal S$-\LaTeX}
\def\BibTeX{{\rmfamily B\kern-.05em
 \textsc{i\kern-.025em b}\kern-.08em
  T\kern-.1667em\lower.7ex\hbox{E}\kern-.125emX}}
\makeatletter
\def\tmpcite#1{\@ifundefined{b@#1}{\textbf{?}}{\csname b@#1\endcsname}}%
\makeatother
\field{D}
\vol{107}
\no{1}
\title[Extraction of Speech Rhythm-Based Speaker Embeddings from Phonemes and Phoneme Duration ]
      {Speech Rhythm-Based Speaker Embeddings Extraction \\from Phonemes and Phoneme Duration \\for Multi-Speaker Speech Synthesis}
\authorlist{%
 \authorentry{Kenichi FUJITA}{m}{labelA}\MembershipNumber{}
 \authorentry{Atsushi ANDO}{m}{labelA}\MembershipNumber{}
 \authorentry{Yusuke IJIMA}{m}{labelA}\MembershipNumber{1411977}
}
\affiliate[labelA]{The author are with the
\EICdepartment{NTT Human Informatics Laboratories}
\EICorganization{NTT Corporation}
\EICaddress{Yokosuka-shi, 239–0847 Japan}
}
\received{2023}{3}{2}
\revised{2023}{8}{6}
\begin{document}
%\linenumbers
%%\linenumberdisplaymath
%%\realpagewiselinenumbers
%%\runninglinenumbers
%%\pagewiselinenumbers
\maketitle

\begin{summary}
This paper proposes a speech rhythm-based method for speaker embeddings to model phoneme duration using a few utterances by the target speaker. Speech rhythm is one of the essential factors among speaker characteristics, along with acoustic features such as F0, for reproducing individual utterances in speech synthesis. A novel feature of the proposed method is the rhythm-based embeddings extracted from phonemes and their durations, which are known to be related to speaking rhythm. They are extracted with a speaker identification model similar to the conventional spectral feature-based one. We conducted three experiments, speaker embeddings generation, speech synthesis with generated embeddings, and embedding space analysis, to evaluate the performance. The proposed method demonstrated a moderate speaker identification performance (15.2\% EER), even with only phonemes and their duration information. The objective and subjective evaluation results demonstrated that the proposed method can synthesize speech with speech rhythm closer to the target speaker than the conventional method. We also visualized the embeddings to evaluate the relationship between the distance of the embeddings and the perceptual similarity. The visualization of the embedding space and the relation analysis between the closeness indicated that the distribution of embeddings reflects the subjective and objective similarity. 
\end{summary}

\section{Introduction}\label{sec:introduction}
Speech rhythm is one of the essential aspects of a human voice. It affects perceptions of speaker similarities along with other acoustic features such as fundamental frequencies. It has been reported that skilled impersonators change their speech style (e.g., speech rhythm, pauses) and F0 to those of target speakers'~\cite{zetterholm2002intonation,zetterholm2003same,gomathi2012analysis}. In these previous studies, the original target speech is compared with the imitated speech from amateur and professional impersonators. They found that even the amateurs changed their speech style.

Many deep neural network (DNN)-based speech synthesis methods have made it possible to reproduce the characteristics of the target speakers~\cite{hojo2018dnn,cooper2020zero,fan2015multi}. These DNN-based speech synthesis methods use auxiliary information in addition to linguistic information to model the speakers' characteristics. Some studies use speaker codes, which are implemented as one-hot vectors corresponding to each speaker, as the auxiliary input~\cite{luong2018scaling,luong2019training}. These conventional methods with speaker codes improve the naturalness of the synthesized speech but are only applicable for speakers that are in the training data. Others use speaker embedding vectors such as the i-vector~\cite{dehak2010front}, d-vector~\cite{heigold2016end}, and x-vector~\cite{snyder2018x} to synthesize speakers unseen in the training data~\cite{wu2015study,doddipatla2017speaker}. For example, Wu {\sl et al.} applied the i-vector~\cite{wu2015study}, and Doddipatla {\sl et al.} applied the d-vector~\cite{doddipatla2017speaker} for speaker adaptation. These studies showed that methods with speaker embedding vectors~\cite{cooper2020zero,jia2018transfer} accurately reproduced individual utterance features, especially in F0 and spectral features.
However, these conventional speech synthesis methods have trouble reproducing speech rhythm. The conventional speaker embeddings trained from spectral features mainly focused on modeling acoustic features (spectral features and F0), which results in ignoring speech rhythm information. It is required to use speech rhythm indicators as the auxiliary features to improve speaker similarities of the synthesized speech.

To this end, we propose a method to extract the speaker embeddings that contain rhythm information. The key idea of the proposed method is to utilize the speaker identification model from rhythm-based features to extract speaker-specific rhythm characteristics. As each speaker has a speaking rhythm, we hypothesize that the embeddings reflecting the speech-rhythm features are extracted from a bottleneck layer of the speaker identification model in the same way with the x-vector. Our previous work~\cite{fujita2021phoneme} demonstrated that speaker identification from phonemes and their duration was sufficient for extracting embeddings, and speech synthesis with the proposed embeddings improves the reproduction quality for the target speaker. In this paper, we analyzed the training data and embedding space, in addition to the experiments conducted in the previous work, for further understanding of the embeddings.

To analyze the training data, we trained the speaker identification model using datasets containing a different number of speakers to reveal the relationship between the quality of embeddings and the diversity in the datasets, as the dataset was fixed in the past research. To verify that the extracted embeddings reflect speech rhythm features, we analyzed the relation between the similarity in the proposed embeddings space and that determined by objective and subjective evaluations. The analysis of embedding space was limited to quantitative analysis by visualization in our previous work. We only showed that the utterances with a similar speaking rhythm seem to be also close in the proposed embedding space.

We conducted objective and subjective evaluations to show that the proposed method extracts rhythm-based embeddings for synthesizing speech with higher similarities to the target speakers. As in our previous work~\cite{fujita2021phoneme}, the speaker identification showed that the proposed method achieved a moderate equal error rate (EER) of 15.2\%, which is sufficient for extracting speech rhythm embeddings. The synthesized speech with the proposed embeddings outperformed that with the conventional embeddings (x-vector) in both objective and subjective evaluations. Additional experiments showed that the model's performance improves with a larger number of speakers in the training dataset. The relation analysis indicated that the similarity in the proposed embeddings was correlated to the objective and subjective similarity of the utterances.

\section{Related work}
\subsection{Multi-speaker speech synthesis}
In multi-speaker speech synthesis, pre-trained and jointly-optimized speaker representations are introduced as auxiliary information. As described in Sect.~\ref{sec:introduction}, some conventional methods use embeddings extracted from the pre-trained speaker encoder. For example, the embeddings extracted from the middle layer of a pre-trained speaker identification model, such as d-vector~\cite{heigold2016end} and x-vector~\cite{snyder2018x}, are used as auxiliary information. Typically, the speaker identification model is trained to identify speakers from acoustic features and does not explicitly focus on speech-rhythm features. However, embeddings reflecting the speech-rhythm features are necessary for higher quality reproduction of the target speakers. Extracting speech rhythm-based embeddings is the goal of our research.

Some studies acquired speech synthesis models containing a speaker characteristic extraction model jointly-trained with speech synthesis, e.g., Deep Voice 2~\cite{NIPS2017_c59b469d} and global style tokens~\cite{wang2018style}. Deep Voice 2 uses trainable embedding vectors for each speaker to condition speech synthesizers. After training, the embedding space which reflects the individual characteristics is automatically acquired in the model. Speech synthesis with global style tokens also includes trainable vectors for extracting individual characteristics in the model. This method demonstrated that each vector reflects an aspects of a speakers' style (e.g., speaking rate, fundamental frequency). However, these latent style variables are entangled, and it is difficult to control the style of the generated speech during inference. Entangled variables also mean the extraction of speech rhythm features is not guaranteed. Therefore, we used a pre-trained speech encoder to explicitly model speech rhythm for multi-speaker speech synthesis.

\subsection{Embedding extraction and speech rhythm modeling}
Our proposed method uses a speaker identification model trained on a sequence of phonemes and their durations to extract rhythm-based embeddings. Conventionally, spectral features have been used as an input to the identification model. Prior studies have also introduced other features related to the speaker's identity as additional information to improve the accuracy and robustness of the identification model. Several studies have reported that speaker identification accuracy can be improved by using pitch features in addition to spectral features~\cite{carey1996robust,miyajima2001speaker}. Wang et al.~\cite{wang2019usage} and Liu et al.~\cite{liu2018speaker} used phoneme information as additional information for speaker identification. They proposed methods for generating speaker embedding vectors from spectrograms by multi-task learning with speaker identification and phoneme identification to extract more text-independent vectors reflecting individual features. However, there have not been studies on using rhythm information for speaker identification.

Another possible approach for speech rhythm modeling is the speaker clustering-based approach (such as~\cite{ijima2015statistical}) by using global speech rhythm-related features, i.e., speaking rate. Although these clustering approaches can take global speech rhythm into account easily, it would be difficult to precisely reproduce the speech rhythm for various speakers, including acted, non-professional, and L2 speakers, using speaker clustering and global speech rhythm-related features only. Therefore, not only global speech rhythm features but local ones would be an important factor for modeling speech rhythm.

Advanced non-autoregressive speech synthesis models, e.g., Fastspeech2~\cite{ren2021fastspeech} and Parallel Tacotron~\cite{elias2021parallel}, include duration prediction to control speaking rhythm in speech synthesis. These architectures model speech rhythm; however, their goal is to model speech rhythm in an utterance conditioned by speaker embeddings. In contrast, our study aims to generate embeddings that capture the speaker's speech rhythm characteristics.

\begin{figure}[tb]
    \centering
    \includegraphics[width=0.9\linewidth]{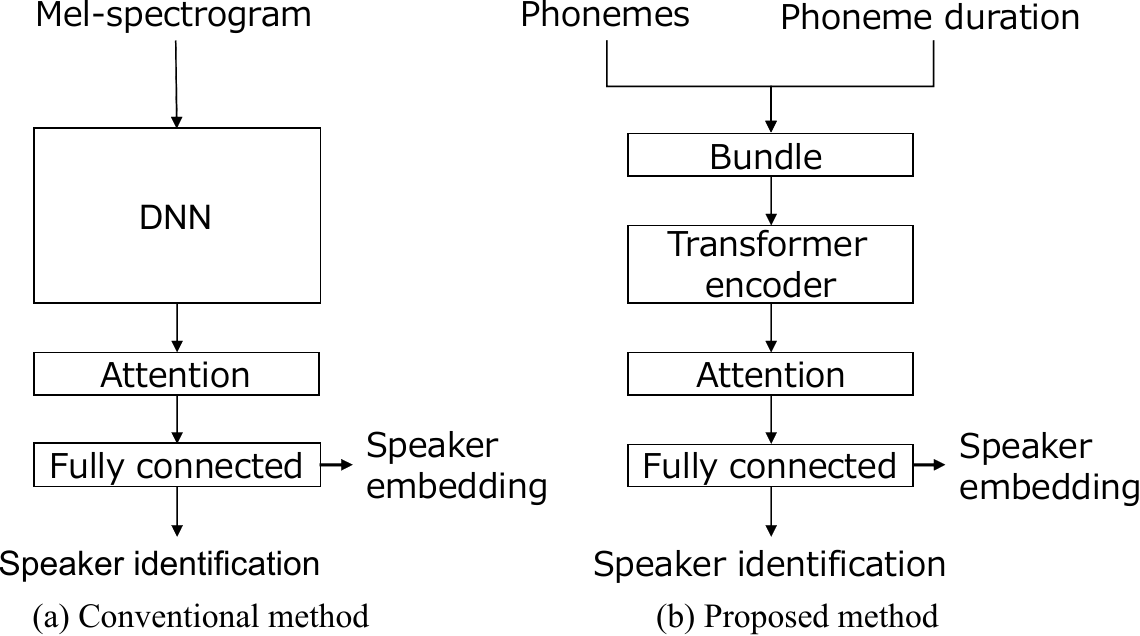}
    \caption{Comparison of conventional and proposed methods.}
    %\ecaption{Conventional and proposed methods}
    \label{fig:methods1}
\end{figure}

\section{Speaker embedding method}\label{sec:propose_method}
\subsection{x-vector}
This section describes the conventional x-vector~\cite{snyder2018x} that is widely used as speaker embedding for multi-speaker speech synthesis. The x-vector is extracted from the middle layer of the DNN model (Fig.~\ref{fig:methods1}(a)) trained to identify speakers from input acoustic features. The training results in a speaker identification model with speaker classification loss (e.g., softmax) or metric learning loss (e.g., angular prototypical~\cite{chung2020in}) applied for the output of the model. After the training with utterances from a large number of speakers, the x-vector is extracted from a bottleneck layer of the DNN, one layer just before the output.

The speaker identification model consists of three main blocks (Fig.~\ref{fig:methods1}(a)). The first is the DNN block that extracts local-level features from the input acoustic features. It consists of multiple time-delay neural networks (TDNNs)~\cite{peddinti2015time}, ResNet~\cite{chung2020in}, or Transformer Encoder layers~\cite{nj2021s}. The original x-vector uses TDNN for the DNN architecture, and subsequent architectures have been used to improve the speaker identification performance. The second block is the Attention block, which aggregates local-level features into whole sentence-level features as a single compact representation of arbitrary length. Originally, statistic pooling was used for this aggregation~\cite{snyder2018x}, but the attention structure has been found to extract features more accurately~\cite{cai2018exploring,rahman2018attention}. Finally, the whole-sentence-level features pass through fully-connected layers.

\subsection{Proposed rhythm-based speaker embeddings}\label{sec:model_structure}
This section introduces the rhythm-based speaker embedding extraction method. We constructed a speaker identification model that leverages phonemes and their durations as input to the model. The model is modified to extract features from the input (Fig.~\ref{fig:methods1}(b)). (1) The bundle block is introduced because the duration of each phoneme is expected to be strongly connected with phonemes and their durations near the target phonemes. (2) The Transformer encoder is used for the DNN block because the speech rhythm feature derives from a longer speech context. (3) Self-attentive pooling is applied to the temporal direction in the Attention block to extract features related to time. We will explain the input features and proposed model structure in the following sections. 

\subsubsection{Input features}
The input to the proposed model is the speech rhythm representing features, i.e., pairs of phonemes and their durations. One-hot vectors expressing phonemes and one-dimension vectors representing phoneme duration are concatenated and used as the input feature to the proposed speaker identification model. Therefore, the input feature is expressed as $\boldsymbol{X}=[\boldsymbol{x}_1, \cdots, \boldsymbol{x}_T]$ where $\boldsymbol{x}_t \in \mathbb{R}^{K+1}$, and $K$ is the number of types of phonemes.

The acoustic feature, which is the input feature of the conventional method, also includes the speech rhythm information. However, the model tends to weigh on other more explicit features included in acoustic features when used as an input feature for the speaker identification model.

\subsubsection{Bundle block}
\begin{figure}[tb]
    \centering
    \includegraphics[width=0.8\linewidth]{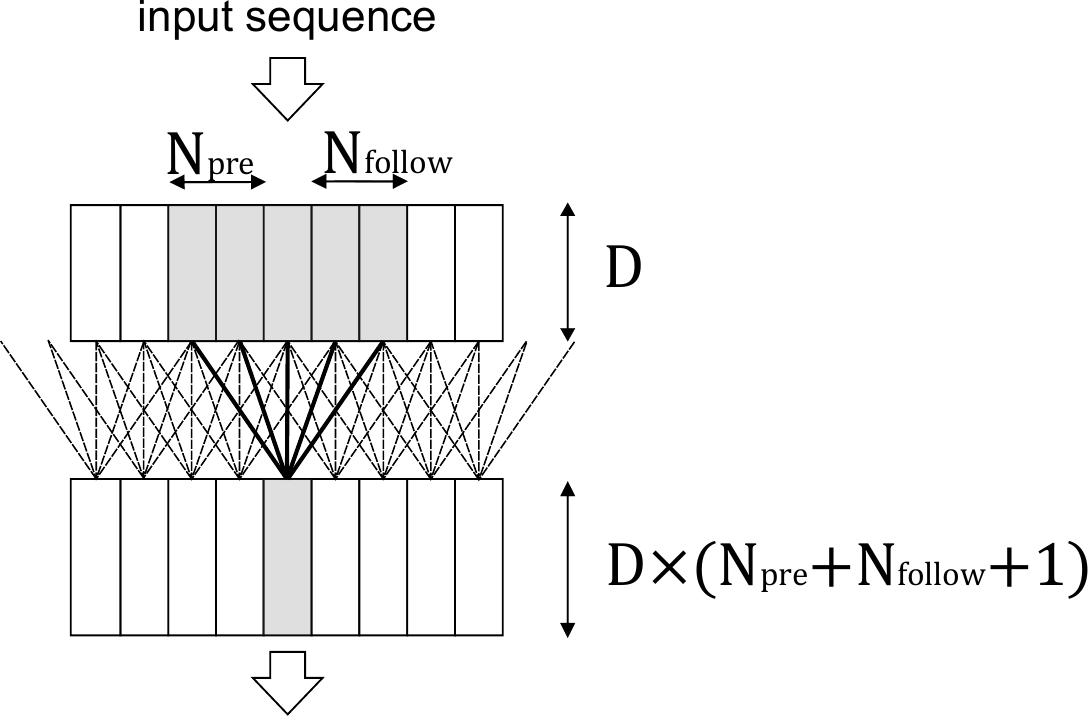}
    \caption{Bundle block.}
    %\ecaption{Conventional and proposed methods}
    \label{fig:bundle}
\end{figure}

The bundle block splices several of the preceding and following input feature vectors ($N_\mathrm{pre}$, $N_\mathrm{follow}$; the number of preceding and following concatenated features, respectively) (Fig.~\ref{fig:bundle}) to extract local features.
Let the input sequence to the block be $\boldsymbol{X}_\mathrm{bundle} = [\boldsymbol{x}_\mathrm{bundle}^1,\dots, \boldsymbol{x}_\mathrm{bundle}^T]$ where $\boldsymbol{x}_\mathrm{bundle}^i \in \mathbb{R}^{D}$. The output sequence $\boldsymbol{O}_\mathrm{bundle} = [\boldsymbol{o}_\mathrm{bundle}^1,\dots, \boldsymbol{o}_\mathrm{bundle}^T]$ is defined by:
\begin{equation}
\begin{split}
    \boldsymbol{o}_\mathrm{bundle}^i&=\mathsf{Concat}(\boldsymbol{x}_\mathrm{bundle}^{i-N_\mathrm{pre}},\dots,
    \\&  \qquad \qquad \boldsymbol{x}_\mathrm{bundle}^i,\dots,\boldsymbol{x}_\mathrm{bundle}^{i+N_\mathrm{follow}})
    \label{eq:bundle}
\end{split}
\end{equation}
where $\boldsymbol{o}_\mathrm{bundle}^i \in \mathbb{R}^{D(N_\mathrm{pre}+N_\mathrm{follow}+1)}$.

\subsubsection{Transformer encoder block}
The Transformer encoder extracts features by looking at the entire input sequence. This feature extraction is suitable for speech rhythm because the speech rhythm feature derives from a long speech context.

The Transformer encoder block~\cite{vaswani2017attention}, a block with positional encodings and several stacked blocks of self-attention and ResNet on top of each other, extracts the features of the series. This encoder has been used in many feature extraction models such as BERT~\cite{kenton2019bert}, GPT-3~\cite{brown2020language}, and HuBERT~\cite{hsu2021hubert}. The block is expressed as: 

\begin{align}
\boldsymbol{H} = \mathsf{TransformerEncoder}(\boldsymbol{X}; \theta)
\end{align}
where $\boldsymbol{H}$ is the output sequence, $\boldsymbol{X}$ is the input sequence, and $\theta$ is the model parameters.

\subsubsection{Attention block}
The temporal direction attention is used for the Attention block to extract the rhythmically characteristic part. The conventional methods use attention in the feature-dimension direction to determine the characteristic frequency in feature extraction from acoustic features.

Let $\boldsymbol{X}_\mathrm{att}=[\boldsymbol{x}_\mathrm{att}^1,\dots,\boldsymbol{x}_\mathrm{att}^T]$ where $\boldsymbol{x}_\mathrm{att}^i \in \mathbb{R}^d$ be the input sequence. Hidden representations $\boldsymbol{h}_i \in \mathbb{R}^M$ ($i=1,...,T$) are calculated as:
\begin{align}
    \boldsymbol{h}_i = \tanh (\boldsymbol{x}_\mathrm{att}^i\boldsymbol{W}+\boldsymbol{b})
\end{align}
Where $\boldsymbol{W} \in \mathbb{R}^{d \times M}$ is the projection matrix.

Then, the output vector $\boldsymbol{O}_\mathrm{att} \in \mathbb{R}^{d}$ is calculated with learnable vector $\boldsymbol{\mu} \in \mathbb{R}^M$ by the weighted average of the frames of $\boldsymbol{X}_\mathrm{att}$ calculated as:

\begin{align}
   \boldsymbol{O}_\mathrm{att} = \sum_{j=1}^T{ \mathsf{softmax}(\boldsymbol{h}_j \boldsymbol{\mu}^\top)\boldsymbol{x}_\mathrm{att}^j}
\end{align}

\section{Experiments on speaker identification}\label{sec:exp}
We conducted speaker identification experiments to verify that the proposed method can capture speaker individuality from pairs of phonemes and their durations.
% The goal of this experiment was to evaluate speaker individuality in pairs of phonemes and their duration through speaker identification. If a pair has the features of each speaker, the speaker identification will show at least a moderate score. Therefore, the proposed method may lead to a lower performance than the conventional methods designed for speaker identification.

\subsection{Dataset}\label{subsubsec:emb_database}
The experiment used an internal Japanese speech database containing 920 speakers. This database consists of several speaker types including professional speakers, i.e., newscasters, narrators, and voice actors, non-professional speakers, and an L2 speaker. Since this dataset was constructed for text-to-speech, the nature is different from that of spontaneous speech. 
\begin{enumerate}
\item Each speaker was instructed to maintain a constant speaking rate and tone during the recording. \item Some speakers recited the same script. 
\item The dataset includes manually labeled phoneme information and phoneme duration information obtained from the phoneme boundary information from manually segmented utterances. 
\end{enumerate}
We prepared four training datasets containing varying numbers of speakers (100, 200, 400, 800) and utterances (73,981, 89,650, 101,208, 126,238), and the validation and test dataset included 8,137 from 60 speakers and 900 utterances from 60 speakers, respectively. The four datasets were used to verify the relation between speaker identification and the diversity of the speakers in the training dataset. For this validation and test dataset, a random trial set of 2,700 same and 2,700 different pairs are generated for calculating equal error rate (EER).

\subsection{Model configurations}\label{subsubsec:emb_conditions}
We trained the conventional x-vector and proposed models to evaluate their speaker identification performance. The structure of the proposed model is described in Sect.~\ref{sec:model_structure}. 
The input features were 56-dim one-hot phoneme vectors and the phoneme duration. The model was a two-layer Transformer encoder that was preceded by the bundle block and followed by the attention block and fully-connected block. Each hidden layer contained 300 hidden nodes. The bundle block contained $N_{pre} = 2$ preceding features and $N_{follow} = 2$ following features. The Transformer encoder consisted of two layers, 64 units, and eight heads. The attention layer aggregates input features into 32-dim feature vectors by using a self-attentive structure with 64 units and eight heads. The fully-connected block output 32-dim bottleneck features. These parameters were experimentally determined. 
We used angular prototypical loss, a metric learning loss, for up to 1000 epochs and evaluated every 10 epoch with EER. The training was stopped when the EER reached the lowest value in the validation set. To evaluate the contribution of phoneme duration to speaker characteristic extraction, we also trained the following two models: the model only trained with one-hot phoneme vectors and that trained with only phoneme duration.

The structure of the conventional model (x-vector) is Fast ResNet-34, which consists of a convolution layer and multi-stage ResNet layers~\cite{chung2020in}. The input acoustic feature was 40-dim mel-spectrograms extracted by mel-spectrogram analysis computed every 10 ms with a 25 ms Hamming window from 16 kHz speech signals. We used self-attentive pooling (SAP)~\cite{cai2018exploring} for the attention layer. As in the proposed method, the dimension of the speaker embedding vector was set to 32, and we used angular prototypical loss.

We call the conventional model ``x-vector'', the proposed model trained with phonemes and their durations ``proposed'', that with only phonemes ``phonemes only,'' and that with only phoneme duration ``duration only.''

\begin{figure}[tb]
    \centering
    \includegraphics[width=0.65\linewidth]{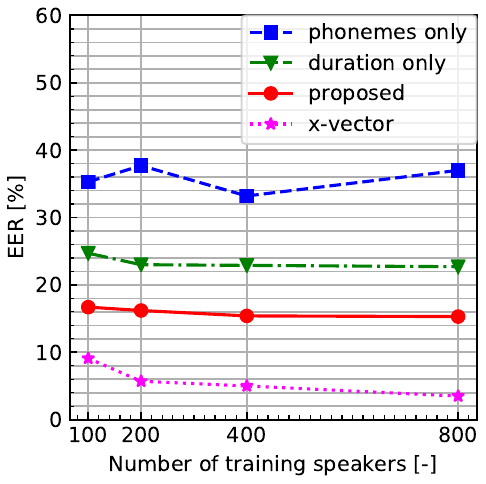}
    \caption{Results of speaker identification.}
    %\ecaption{The structure of the proposed model}
    \label{fig:mineer}
\end{figure}

\subsection{Evaluation of speaker identification performance}\label{subsec:speakerrecog_result}
Figure~\ref{fig:mineer} shows the speaker identification performance of each model. The chance rate of the test data was 50.0\%. The EER of the proposed model was about 17\% in every case and achieved 15.2\% for 800 speakers. This result indicates that we can extract speaker characteristics to some extent from only phonemes and their durations without using acoustic features such as spectral features and F0. Furthermore, the performance of the proposed model exceeded that of duration-only and phonemes-only models in every case. This result shows using both phonemes and their durations leads to improve speaker characteristic extraction. Note that the EER score of phoneme only is ideally 50.0\% because identifying a speaker only from text is difficult. The EER score of phoneme only was lower than 50\%, suggesting that some speech in the dataset was phonetically biased. One reason is that the dataset is for speech synthesis, and some of the speakers uttered the same scripts.

\begin{figure*}[tb]
  \begin{minipage}[b]{0.33\linewidth}
    \centering
    \includegraphics[width=1.05\linewidth]{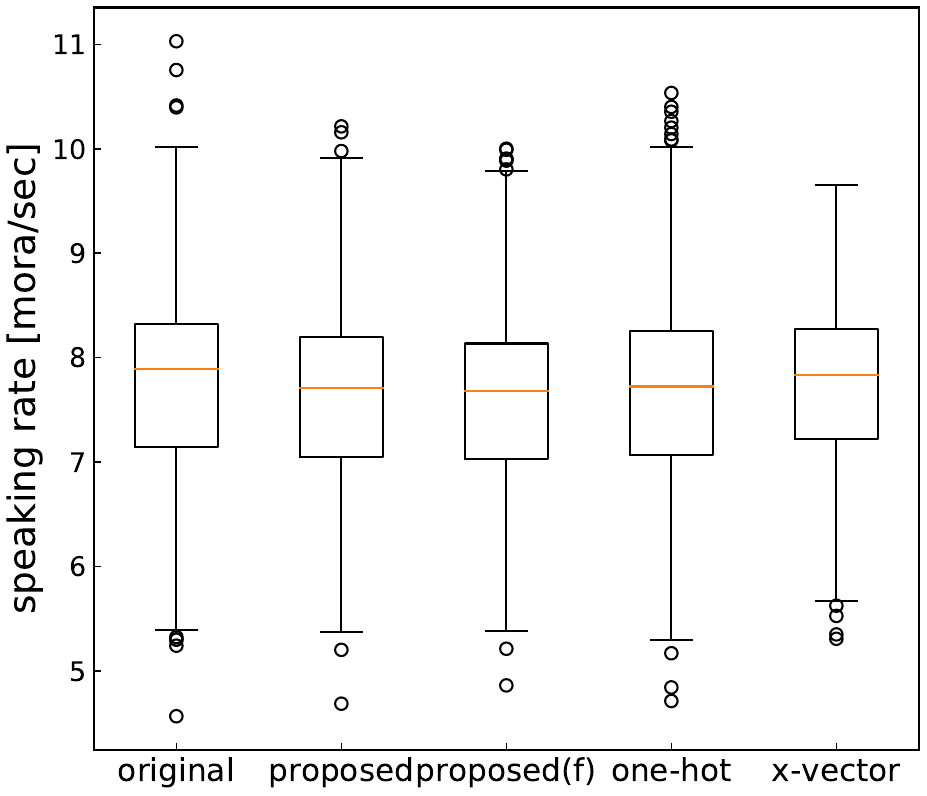}
    \subcaption{Speaking rates.}
    \label{fig:boxplots2}
  \end{minipage}
  \begin{minipage}[b]{0.33\linewidth}
    \centering
    \includegraphics[width=0.85\linewidth]{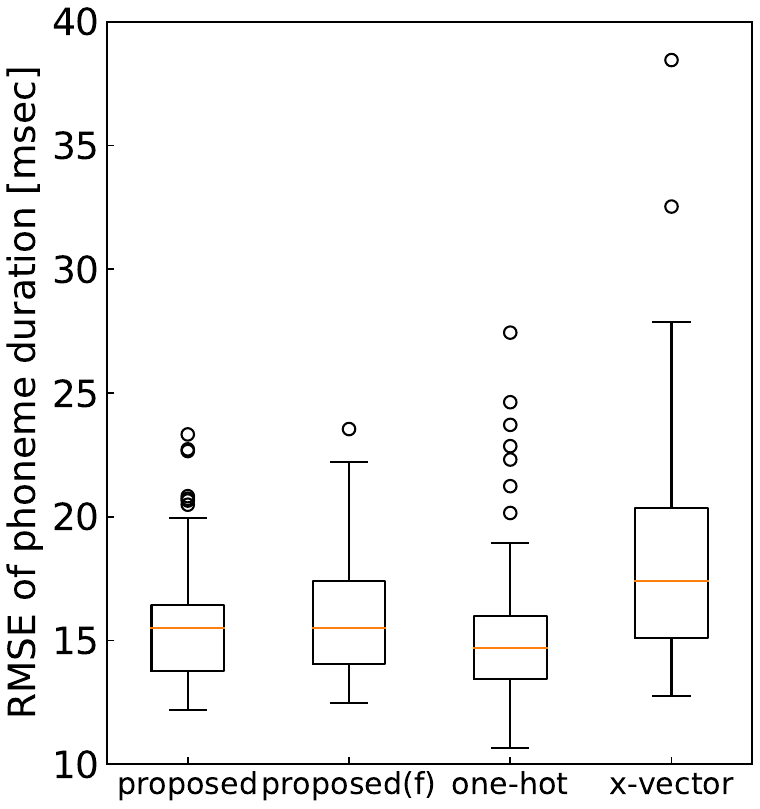}
    \subcaption{Root mean square errors.}
    \label{fig:boxplots}
  \end{minipage}
  \begin{minipage}[b]{0.33\linewidth}
    \centering
    \includegraphics[width=0.9\linewidth]{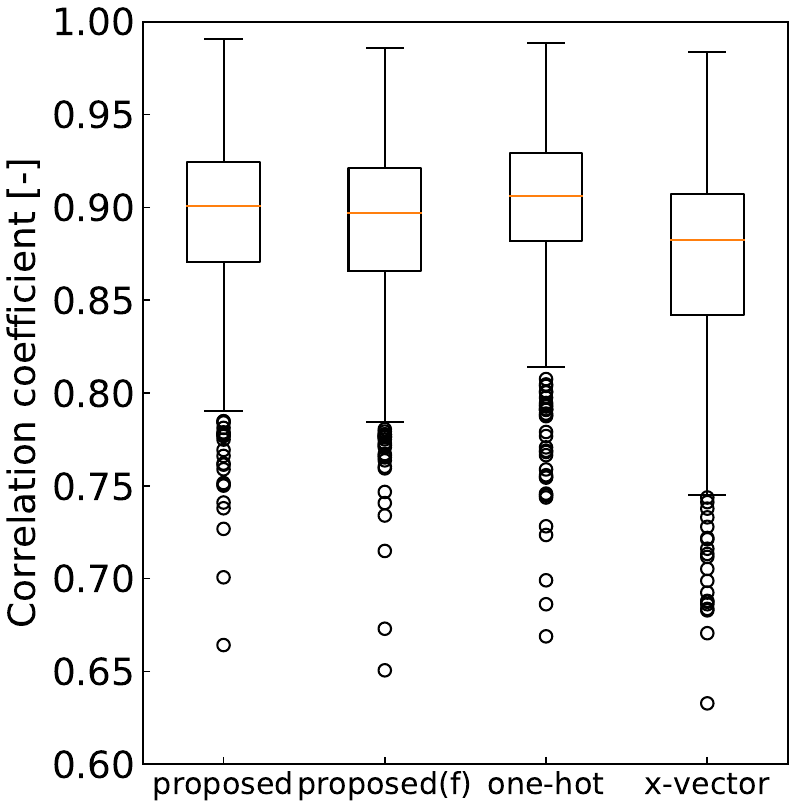}
    \subcaption{Correlation coefficients.}
    \label{fig:boxplots_rel}
  \end{minipage}
  \caption{Results of objective evaluations in phoneme duration prediction. ``proposed(f)'' indicates the proposed method using automatically estimated phoneme durations.}
  \label{fig:boxplot_figs}
\end{figure*}

Compared with the performance of the x-vector, that of the proposed model was lower in every case. This indicates that the x-vector extracts more dominant speaker characteristics, i.e., acoustic features, than speaking rhythm. The proposed model extracts speaker characteristics to some extent without using the dominant features.

\section{Experiment for phoneme duration prediction}\label{subdec:condition}
We conducted objective and subjective evaluations to investigate the performance of phoneme duration prediction using the proposed speaker embeddings. 
%If the speaker identification model extracts individual speech rhythm, the phoneme duration prediction with the proposed embeddings is expected to reproduce the speech rhythm of the target speakers. 

\begin{table}[tb]
  \caption{Training dataset for phoneme duration prediction.}
  \label{table:exp2_cond}
  \centering
  \begin{tabular}{l|ccc}\hline
             &proposed  &one-hot &x-vector\\
    \hline\hline
    Train-A &\checkmark & \checkmark &\checkmark \\
    Train-B &$\times$ &\checkmark &$\times$\\
    \hline
  \end{tabular}
\end{table}

\subsection{Experimental conditions}
As the phoneme duration model, we adopted an explicit duration prediction model in the same manner as DNN-based speech synthesis~\cite{ze2013statistical} and a non-autoregressive speech synthesis model, e.g., FastSpeech2~\cite{ren2021fastspeech}. This is because directly evaluating the predicted duration with a sequence-to-sequence model, e.g., Tacotron2~\cite{shen2018natural}, is difficult. The model consists of six Transformer encoder blocks. Each block has a dimension of 64 with eight heads. The input is a 303-dimensional linguistic vector and the speaker embedding vector. The output is the one-dimensional duration of each phoneme. For training, we used mean squared error (MSE) loss. The model was trained up to 1000 epochs until the MSE of the validation set yielded the lowest score.

We trained the model using the same database described in Sect.~\ref{subsubsec:emb_database} with the three types of vectors, i.e., the proposed embedding vector, one-hot vector, and x-vector. These vectors were extracted from the model trained with 800 speakers. The proposed and x-vector methods were trained on an open-speaker condition, whereas the one-hot method was trained with a closed-speaker case. We trained the one-hot model as an ideal case where retraining the duration prediction model with the target speaker's utterances is allowed. Furthermore, in evaluation, we compared the performance of the proposed method with two types of phoneme durations for its input; the manually annotated and the automatically estimated by forced phoneme alignment. The aim is to confirm the robustness for the automatically annotated phoneme durations, since manually annotated phoneme durations are often unavailable in the practical use of speech synthesis. To obtain automatically estimated phoneme durations, we applied DNN-HMM-based forced alignment, trained with the in-house dataset, to the test dataset. The frame shift was 5 ms. The root mean square error, and median and 95 percentile of the alignment error for the test dataset were 8.57 ms, 5.00 ms, and 15.00 ms respectively.

The database was separated into three groups (Train-A, Train-B, and Test) to create open-speaker and closed-speaker datasets. The details of each dataset are as follows. Train-A consisted of 101,208 utterances by 800 speakers, the same speakers used in the speaker identification training. Train-B and Test were from the 60 test speakers. Train-B, which is training data for the test-data speakers, consisted of 300 utterances, five from each of the 60 test-data speakers. This is to evaluate the model trained with a few utterances. Test was 600 utterances comprising ten from each of the 60 test-data speakers, not including the utterances in Train-B. The open-speaker condition was trained with Train-A, and the closed-speaker condition was trained with both Train-A and Train-B as shown in Table~\ref{table:exp2_cond}. As the embedding vector of each speaker, we used the average values from the embedding vectors of all utterances obtained from Train-A and Train-B, i.e., the embedding vectors of the speakers in Train-B were calculated from five utterances included in Train-B.

To evaluate the difference in phoneme duration predicted by each method, the same DNN acoustic model was used for speech parameter generation. We used two linear layers and two unidirectional LSTM layers with 512 units per layer. We used ReLU as the activation function. The acoustic features were generated from each predicted phoneme duration with multi-speaker DNN-based speech synthesis using one-hot speaker code~\cite{hojo2018dnn}. The speech synthesis model generated 80-dim mel-spectrogram and the acoustic features were converted into speech waveforms using a neural waveform generation method, HiFi-GAN~\cite{kong2020hifi}.

% \begin{figure}[tb]
%     \centering
%     \includegraphics[width=0.55\linewidth]{./pictures/boxplot_spkrate_utt.eps}
%     %\vspace{-6mm}
%     \caption{Speaking rate for each condition.}
%     %\ecaption{The plot of speaker embedding vectors (56d)}
%     \label{fig:boxplots2}
% \end{figure}
% \begin{figure}[tb]
%     \centering
%     \includegraphics[width=0.55\linewidth]{./pictures/boxplot_utt.eps}
%     %\vspace{-6mm}
%     \caption{RMSE of phoneme duration for each condition.}
%     %\ecaption{The plot of speaker embedding vectors (56d)}
%     \label{fig:boxplots}
% \end{figure}

% \begin{figure}[tb]
%     \centering
%     \includegraphics[width=0.55\linewidth]{./pictures/boxplot_rel.eps}
%     %\vspace{-6mm}
%     \caption{Correlation coefficient for each condition.}
%     %\ecaption{The plot of speaker embedding vectors (56d)}
%     \label{fig:boxplots_rel}
% \end{figure}

\subsection{Objective evaluation}\label{subsec:pdm_objective_eval}
To evaluate the effectiveness of the proposed method, we first compared the distribution of speaking rates obtained from each method. Figure~\ref{fig:boxplot_figs}(\subref{fig:boxplots2}) shows a box plot of the speaking rate from the 600 utterances (10 utterances from each of 60 speakers) in the test dataset. As shown, original speech had a wide distribution, and that of the proposed and one-hot methods had a similar tendency. In contrast, that of the x-vector was significantly narrower than the other three distributions. The analysis of variance showed that the p-value between the original and proposed, original and one-hot, and original and x-vector were 0.32, 0.68, and $2.50\times10^{-7}$, respectively.

Figures~\ref{fig:boxplot_figs}(\subref{fig:boxplots}) and \ref{fig:boxplot_figs}(\subref{fig:boxplots_rel}) also show a box plot of the root mean square error (RMSE) of phoneme durations for each method and the correlation coefficient of phoneme durations between the predicted and original durations. 
The average RMSEs and correlation coefficient of the proposed and one-hot were almost the same, even though the 60 test-data speakers were open speakers for the proposed model and closed speakers for the one-hot. 
In other words, without the retraining using the target speaker, the proposed model can predict the phoneme durations of the open speaker as accurately as the one-hot model trained with open target speakers' utterances. 
This is an advantage of the proposed method because retraining for speakers not in the training data requires data for adequate adaptation as well as computational time and resources. We can also verify that x-vector shows higher RMSE and lower correlation coefficient than that of the proposed method. This indicates that the proposed speaker-embedding vector can capture speech rhythm more accurately than the x-vector.

As for the performance by using automatically estimated phoneme durations, we can see that a comparable performance is obtained compared with that by using manually annotated ones. This indicates that our proposed method is robust enough for practical use of speech synthesis even though by using automatically estimated phoneme durations.

\begin{figure}[tb]
    %\begin{tabular}{cc}
    %1st figure
    \begin{minipage}[t]{1.0\hsize}
        \centering
        \includegraphics[width=0.7\linewidth]{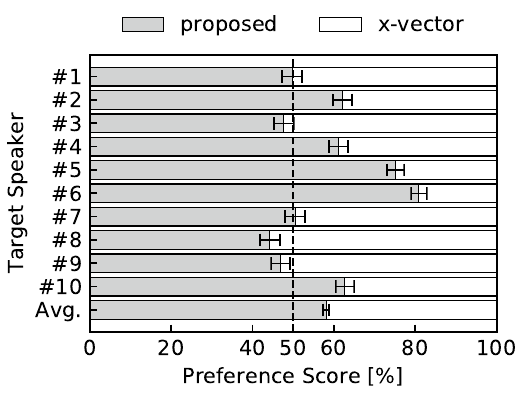}
        \vspace{-3mm}
        \subcaption{Proposed vs x-vector.}
        \label{fig:abx_similar_conv}
    \end{minipage} \\
    %2nd figure
    \begin{minipage}[t]{1.0\hsize}
        \centering
        \includegraphics[width=0.7\linewidth]{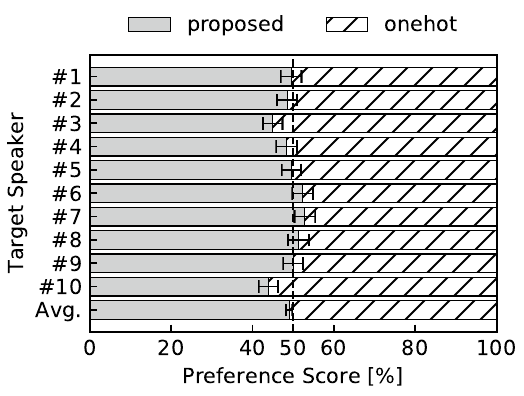}
         \vspace{-3mm}
        \subcaption{Proposed vs one-hot.}
        \label{fig:abx_similar_one-hot}
    \end{minipage}
    \begin{minipage}[t]{1.0\hsize}
      \centering
      \includegraphics[width=0.7\linewidth]{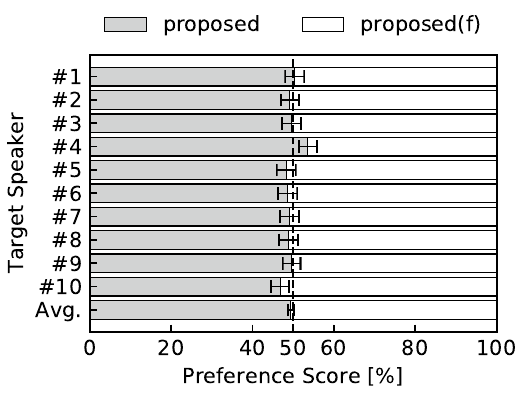}
       \vspace{-3mm}
      \subcaption{Proposed vs proposed(f).}
      \label{fig:abx_similar_pro-prof}
    \end{minipage}
    \vspace{-3mm}
    \caption{Preference score from subjective evaluation in similarity. Error bars show 95\% confidence intervals. ``proposed(f)'' indicates the proposed method using automatically estimated phoneme durations.}
    \label{fig:abx}
\end{figure}

\subsection{Subjective evaluation}
\subsubsection{Subjective evaluation on speaker similarity}\label{subsubsec:similar}
To evaluate the performance of the proposed method, we conducted XAB listening tests on speaker similarity. All permutations of synthetic speech pairs were presented in two orders (XAB and XBA) to eliminate bias in the order of stimuli. To evaluate only the difference in phoneme duration, the reference speech was the synthetic speech from the same DNN acoustic model with the original phoneme duration of the target speaker. The experiment was conducted on a crowdsourcing platform with 400 participants, each evaluating 20 pairs. Each participant was presented with synthesized speech samples and then asked which sample was similar to the reference speech. Ten utterances uttered by each of 10 speakers, were used for the evaluation. The 10 speakers were selected from the 60 test-data speakers sorted by RMSE (Fig.~\ref{fig:boxplot_figs}(\subref{fig:boxplots})) at equal intervals.

\begin{figure}[tb]
    %\begin{tabular}{cc}
    %1st figure
    \begin{minipage}[t]{1.0\hsize}
        \centering
        \includegraphics[width=0.7\linewidth]{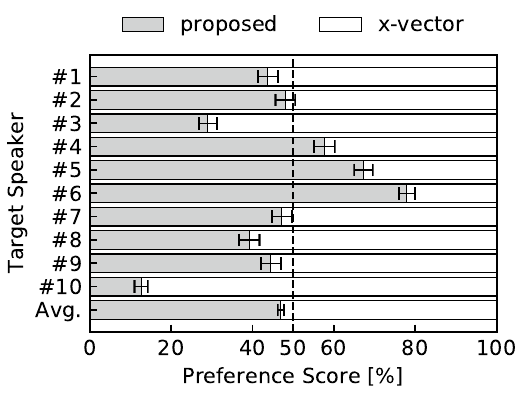}
        \vspace{-3mm}
        \subcaption{Proposed vs x-vector.}
        \label{fig:ab_natural_conv}
    \end{minipage} \\
    %2nd figure
    \begin{minipage}[t]{1.0\hsize}
        \centering
        \includegraphics[width=0.7\linewidth]{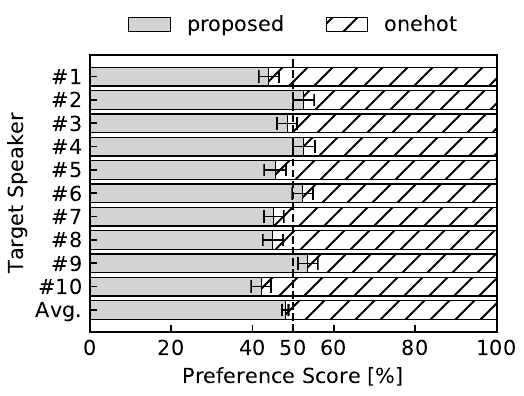}
         \vspace{-3mm}
        \subcaption{Proposed vs onehot.}
        \label{fig:ab_natural_onehot}
    \end{minipage}
    \begin{minipage}[t]{1.0\hsize}
      \centering
      \includegraphics[width=0.7\linewidth]{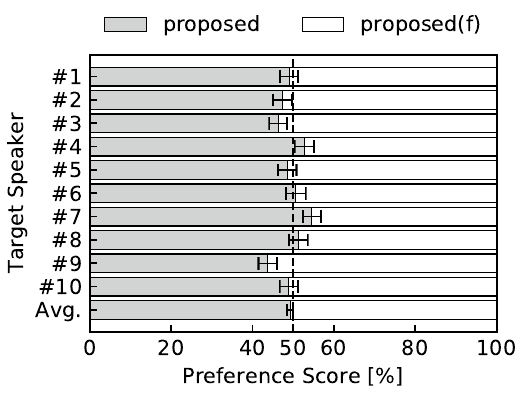}
       \vspace{-3mm}
      \subcaption{Proposed vs proposed(f).}
      \label{fig:ab_natural_pro_prof}
    \end{minipage}
    \vspace{-3mm}
    \caption{Preference score from subjective evaluation in naturalness. Error bars show 95\% confidence intervals. ``proposed(f)'' indicates the proposed method using automatically estimated phoneme durations.}
    \label{fig:ab}
\end{figure}

Figure~\ref{fig:abx} shows the preference scores for each target speaker. The proposed method scored higher than x-vector and was comparable to one-hot vector. This indicates that the proposed method can synthesize speech closer to that speech rhythm of the target speaker than the spectral feature-based x-vector. The proposed method also achieved almost the same quality as that of the one-hot model, even though retraining is not required in the proposed method. Furthermore, the proposed method using automatically estimated phoneme durations showed comparable performance with that using manualy annotated phoneme durations.

\begin{figure*}[tb]
    \begin{minipage}[t]{1.0\hsize}
        \centering
        \includegraphics[width=0.85\linewidth]{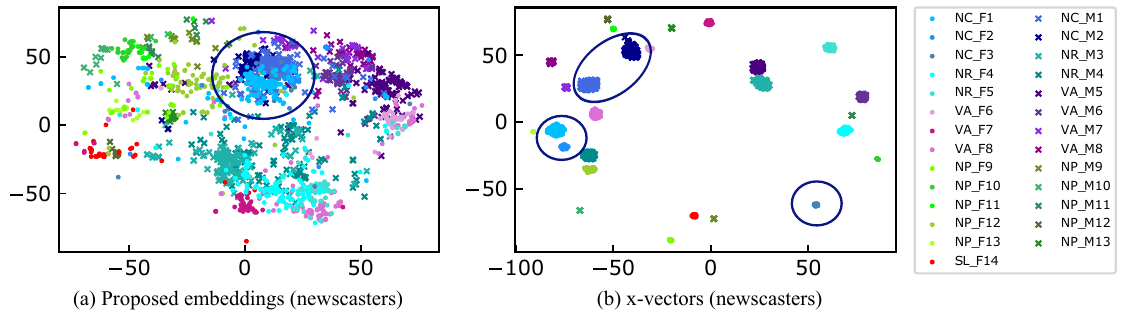}
    \end{minipage} \\
    \begin{minipage}[t]{1.0\hsize}
        \centering
        \includegraphics[width=0.85\linewidth]{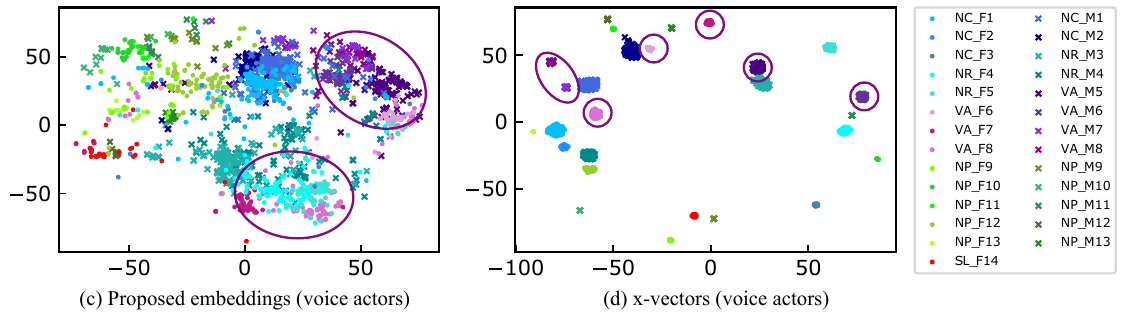}
    \end{minipage} \\ 
    \begin{minipage}[t]{1.0\hsize}
        \centering
        \includegraphics[width=0.85\linewidth]{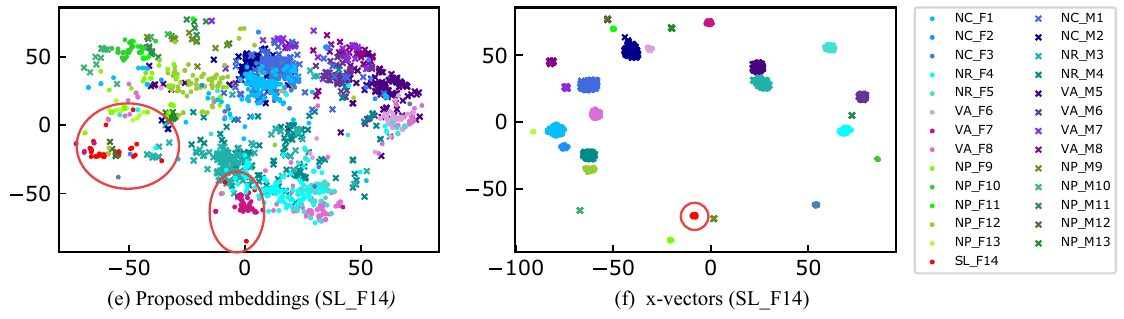}
    \end{minipage} \\ 
    \vspace{-2mm}
    \caption{Distribution of speaker embedding vectors using proposed method. Blue, purple, and red circles indicate newscasters, voice actors, and an L2 speaker, respectively. Female speakers are represented by ``F'' and circles and male speakers ``M'' and crosses. NP: non-professional speakers, NR: narrator (reading style), VA: voice actor (acting as characters), NC: newscasters (reporting style), and SL: L2 speaker.}
    %\ecaption{The plot of speaker embedding vectors (56d)}
    \label{fig:plots}
    \vspace{-5mm}
\end{figure*}

\subsubsection{Subjective evaluation on naturalness}
To evaluate the naturalness of the generated speech, we conducted AB listening tests. All permutations of synthetic speech pairs were presented in two orders (AB and BA). The experiment was conducted on the same crowdsourcing platform with 400 different participants, and each evaluated 20 pairs. Each participant was presented with synthesized speech samples and then asked which sample was more natural. We use the same test data as in Sect.~\ref{subsubsec:similar}.

Figure~\ref{fig:ab} shows the preference scores for each target speaker. The average preference scores show that the naturalness of the proposed method was almost equal to that of x-vector, the one-hot method, and the proposed method with automatically estimated phoneme durations. Likewise, the preference score between proposed method and one-hot had almost the same similarity and naturalness scores for each target speakers. However, the scores between proposed method and x-vector show different tendency. For example, the proposed method yielded lower naturalness scores for speakers \#3 and \#10 than their similarity scores. To investigate this issue, we analyzed the relationship between the naturalness scores and the speaking rates of each target speaker.
Table~\ref{table:spkrate_naturalness} shows the average speaking rate of each speaker's original and generated speech. The bold values show the preferred one in  Fig.~~\ref{fig:ab}(\subref{fig:ab_natural_conv}). Some speakers with low speaking rates (\#3, \#8, \#9 and \#10) had low naturalness scores, even though the proposed method can synthesize speech with the speech rhythm close to that of the original.
Because the original speaking rate of these speakers was much slower than the Japanese average speaking rate (about 8 mora/sec~\cite{han1994acoustic}), the participants may have perceived the speech as unnatural.

\begin{table}[tb]
  \caption{Speaking rate of each generated speaker in mora/sec. Bold indicates the preferred condition of each speaker in Fig.~\ref{fig:ab}(\subref{fig:ab_natural_conv}).}
  \label{table:spkrate_naturalness}
  \centering
  \begin{tabular}{l|ccc}\hline
             &original &proposed  &x-vector\\
    \hline\hline
    \#1& 8.32&8.43  &\bf{8.96} \\
    \#2& 9.33&9.81  &8.44 \\    
    \#3& 7.14&6.97  &\bf{7.63} \\ 
    \#4& 7.30&\bf{7.11}  &6.67 \\     
    \#5& 8.53&\bf{8.51}  &7.52 \\
    \#6& 8.35&\bf{8.33}  &8.19 \\
    \#7& 8.31&7.70  &7.87 \\                 
    \#8& 6.93&6.55  &\bf{7.25} \\    
    \#9& 6.21&6.29  &\bf{6.66} \\
    \#10& 5.87&5.52  &\bf{6.77} \\    

    \hline
  \end{tabular}
\end{table}

\section{Analysis for embedding space}
%Our study aimed to extract speaker embedding vectors capturing speech-rhythm features for better speech generation. The experiments showed a better performance of the proposed method in target speakers' speech reproduction. 
Finally, we further investigate the nature of the extracted embedding to confirm the extracted embeddings reflect speech rhythm. We conducted the following experiments: visualization of embedding vectors, and relation analysis between cosine similarity in embeddings and speech-rhythm similarity. In the relation analysis, we measure the similarity between speakers by subjective evaluations as well as objective metrics because the weak relation to the subjective similarity was pointed out as the problem of the conventional methods~\cite{saito2021perceptual}. 

\subsection{Visualization of speaker embedding vector}\label{sec:speaker_emb}\label{sec:visualize}
We visualized the extracted speaker embedding vectors from the model trained with 800 speakers with t-SNE~\cite{van2008visualizing} to evaluate their spatial distribution. Figures~\ref{fig:plots} show the distributions of the proposed embedding vectors and the x-vector. 

Distributions from the proposed method (Figures~\ref{fig:plots}(a)(c) and (e)) show that speakers with close speech rhythms existed in a closer area in the space. For example, some clusters (newscasters (NC\_F1, NC\_F2, NC\_F3, NC\_M1, NC\_M2) (Fig.~\ref{fig:plots}(a)) and voice actors playing character (VA\_F6, VA\_F7, VA\_F8, VA\_M5, VA\_M6, VA\_M7, VA\_M8) (Fig.~\ref{fig:plots}(c))), whose speech rhythms were close to each other, were constructed in the space (blue and purple circles in the figures). In addition, the L2 speaker SL\_F14, whose utterance was unsteady and whose rhythm changed from utterance to utterance, was widely distributed (area surrounded by red circles in the figure).

The distributions of the x-vector (Figs.~\ref{fig:plots}(b), (d), and (f)), which are based on spectral features, show different tendencies from those of the proposed model. For example, the newscasters and voice actors were not located close to each other in Figs.~\ref{fig:plots}(b) and (d). In addition, SL\_F14, i.e., utterances by the L2 speaker, were not spread out widely in the space in Fig.~\ref{fig:plots}(f) but were spread similarly to the other speakers. These results indicate that the proposed speech rhythm-based vector extracts features that are not reflected enough in the x-vector resulting in more accurate phoneme duration modeling and similar speech generation as described in Sect.~\ref{subdec:condition}.

\begin{figure}[tb]
\centering
\subfloat[Scatter plots of cosine similarity between x-vectors and correlation coefficient between phoneme duration. ($R=0.27$) ]{\includegraphics[width=0.85\linewidth]{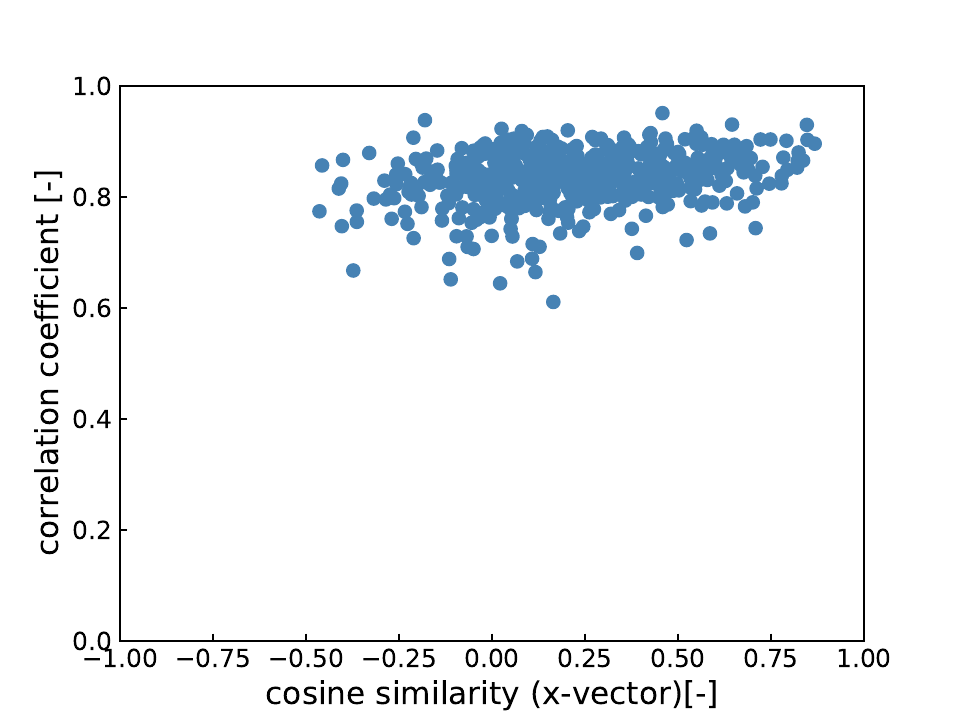}
\label{fig:similar800x_obj}}
\\
\subfloat[Scatter plots of cosine similarity between proposed embeddings and correlation coefficient between phoneme duration. ($R=0.46$)]{\includegraphics[width=0.85\linewidth]{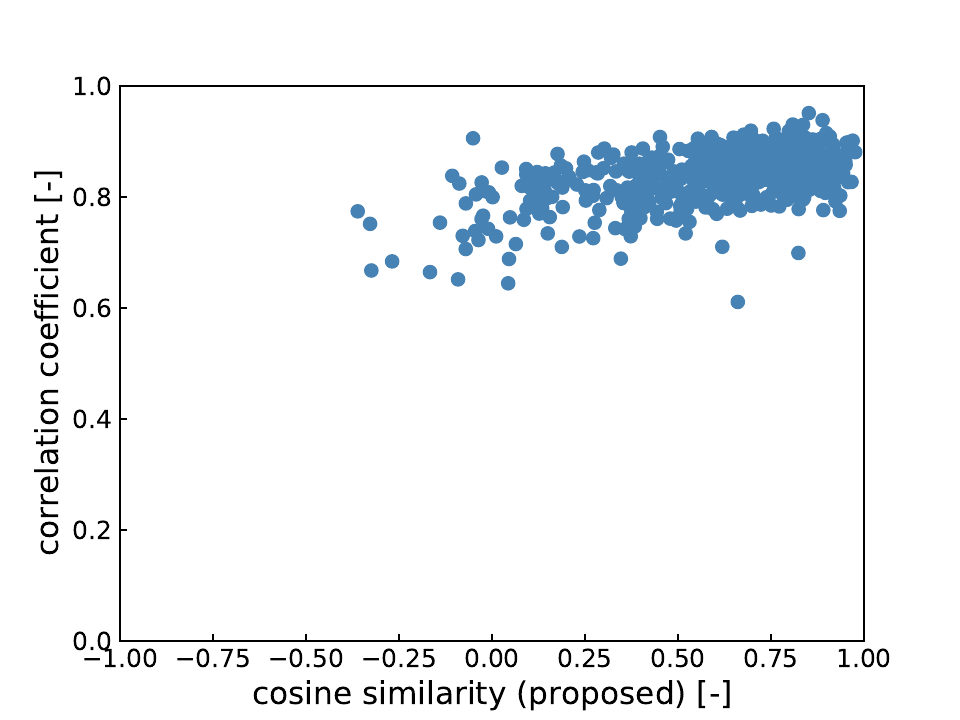}
\label{fig:similar800_obj}}
\caption{Scatter plots of similarity in embeddings and objective similarity.  }
\label{fig:similar_obj}
\end{figure}

\begin{figure}[tb]
\centering
\subfloat[Scatter plots of cosine similarity between x-vectors and subjective similarity. ($R=0.42, MIC=0.31$)]{\includegraphics[width=0.85\linewidth]{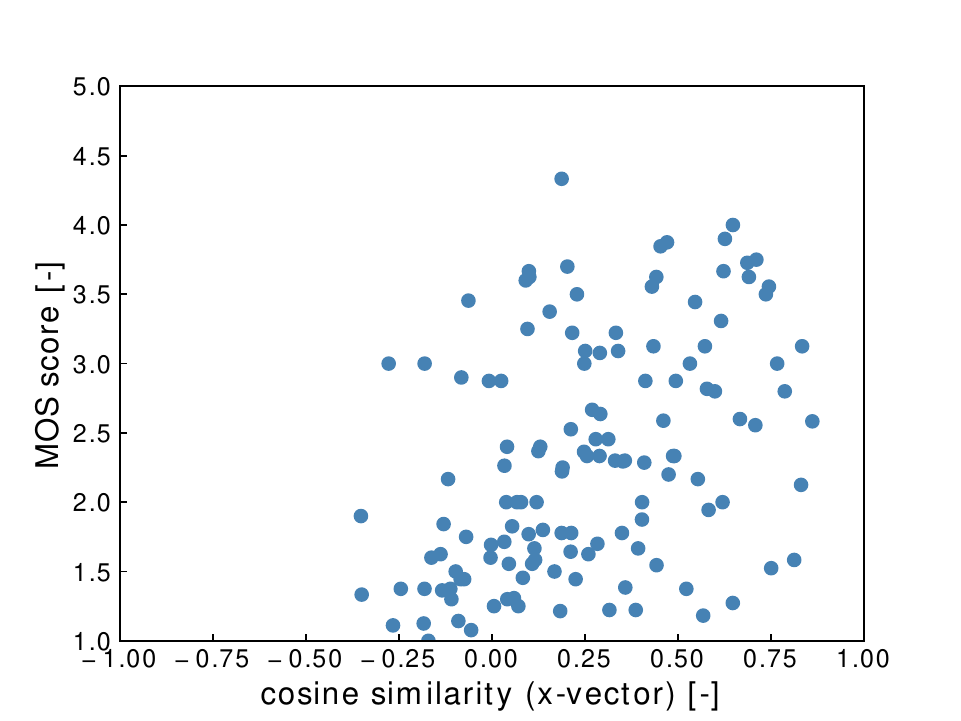}
\label{fig:similar800x_sbj}}
\\
\subfloat[Scatter plots of cosine similarity between proposed embeddings and subjective similarity. ($R=0.46, MIC=0.42$)]{\includegraphics[width=0.85\linewidth]{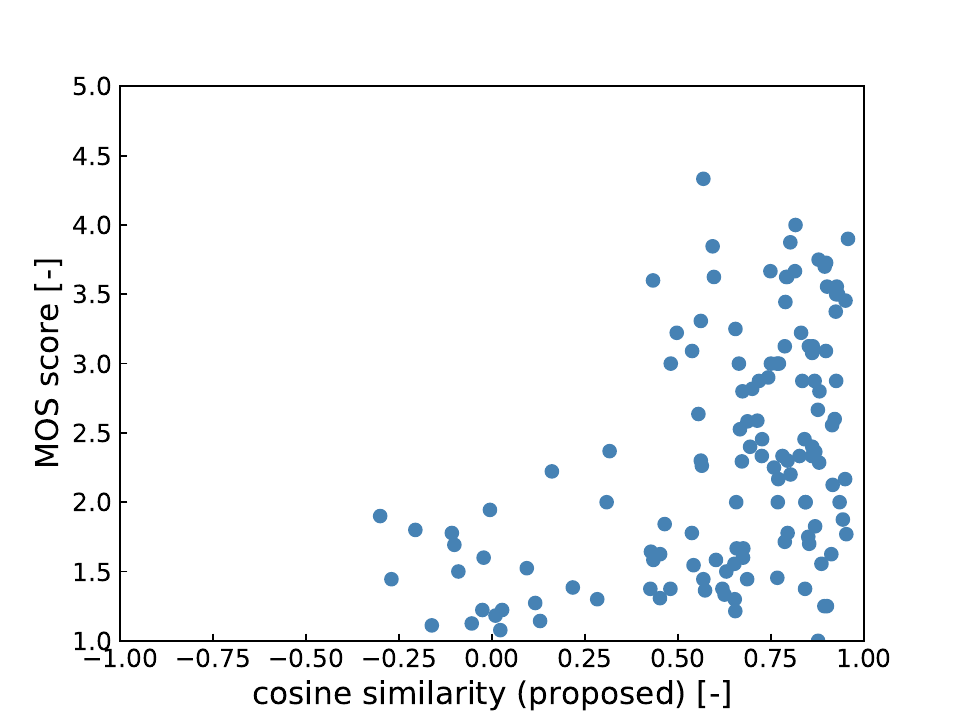}
\label{fig:similar800_sbj}}
\caption{Scatter plots of similarity in embeddings and subjective similarity.}
\label{fig:similar_sbj}
\end{figure}

\subsection{Relation between embedding vectors and speech rhythm}
\subsubsection{Analysis by objective evaluation}\label{subsub:emb_rhythm_obj}
To evaluate the relation between the extracted embeddings and speech rhythm, we first analyzed the relation between the extracted embeddings and the objectively measured similarity. We used the correlation coefficient of phoneme durations between pairs of utterances with the same phoneme sequences as a metric indicating the objective similarity in speaking rhythm.

In this experiment, 14 female and 11 male speakers who uttered the same script were picked from the test dataset. Then, we paired all speakers in the same gender groups. Four utterances per speaker were chosen for the test data. For every utterance pair, the cosine similarity between extracted embeddings from the utterances, and the correlation coefficient of phoneme durations were obtained.
However, phoneme sequences are not necessarily the same between evaluated speakers because the location of silent pause is different in some utterances. To exclude the effect of these different phoneme sequences, we exclude utterances containing different phoneme sequences that differed from the test dataset. 

Figures~\ref{fig:similar_obj}(\subref{fig:similar800x_obj}) and (\subref{fig:similar800_obj}) show the cosine similarities of embedding vectors and correlation coefficients of phoneme duration. 
From Fig.~\ref{fig:similar_obj}(\subref{fig:similar800_obj}), since utterance pairs with higher cosine similarity in the proposed embeddings also have higher cosine similarity in phoneme durations, the proposed method can capture the similarity of speech rhythm to some extent (correlation coefficient of 0.46). In contrast, the x-vector (Fig.~\ref{fig:similar_obj}(\subref{fig:similar800x_obj})) cannot capture the tendency compared with the proposed method (correlation coefficient of 0.27).

These results demonstrated that the similarity in the embedding space is highly related to the speech rhythm similarity, and the proposed method achieved better speech rhythm modeling than conventional x-vector.

\subsubsection{Analysis by subjective evaluation}
Lastly, we evaluate the relationship between the extracted embedding spaces and subjective similarities.

To obtain the subjective similarity, we conducted a subjective evaluation regarding speaker similarity. In the subjective evaluation, we used the same speakers described in the previous section, i.e., 14 female and 11 male speakers. Five non-parallel original utterances (not the synthesized speech) per speaker were used to obtain text-independent subjective similarity among the speakers in a similar manner in~\cite{saito2021perceptual}. Each participant listened to the utterance pairs and rated their similarity on a scale ranging from 5 (very similar) to 1 (very dissimilar). The pairs were presented in two orders (AB and BA) to eliminate bias in the order of stimuli. At least eight participants evaluated each utterance pair. The participants were recruited by crowdsourcing. The subjective similarity between all speakers was scored as the average score of all utterances from the pair. Note that the participants score the subjective similarity with the similarity in acoustic features in addition to the speech rhythm similarity.

Figures~\ref{fig:similar_sbj}(\subref{fig:similar800x_sbj}) and (\subref{fig:similar800_sbj}) show the obtained subjective similarities and the cosine similarities of embedding vectors obtained from each method. The embedding vector of each speaker was the average embedding vector of five utterances extracted from each speaker. For the conventional method (x-vector) described in Fig.~\ref{fig:similar_sbj}(\subref{fig:similar800x_sbj}), speaker pairs with higher cosine similarity did not necessarily have higher subjective scores. Conversely, with the proposed method (Fig.~\ref{fig:similar_sbj}(\subref{fig:similar800_sbj})) most speaker pairs with lower cosine similarity have lower subjective similarity. Additionally, most speaker pairs with higher subjective similarity have higher cosine similarity. 
To quantitatively analyze these distributions, we also obtained the maximal information coefficient (MIC)~\cite{reshef2011detecting} between the subjective similarities and cosine similarities of each method. The MIC makes it possible to detect nonlinear associations that cannot be detected by using the correlation coefficient. Furthermore, the MIC has a property similar to that of the correlation coefficient. That is, the MIC value ranges from 0 to 1, and two variables with a strong association have a value closer to 1. Among the MIC values of each method, the proposed method's was higher value than that of the x-vector. This indicates that the proposed embeddings are related to subjective similarities.

As shown from these analyses, speaking rhythm as well as acoustic features are essential factors to achieving synthesized speech with higher subjective similarity. This is consistent with research on voice mimicry ~\cite{zetterholm2002intonation,zetterholm2003same,gomathi2012analysis}, indicating that subjectively similar utterances should have a similar speaking rhythm as well as acoustic features.

\section{Conclusion}\label{sec:conclusion}

We proposed a method of extracting speaker embedding vectors that capture speech-rhythm features using phonemes and their durations as input. Conventionally, speaker embedding vectors have mainly been based on acoustic features and not explicitly focused on speech rhythm, which is an important factor in speaker characteristics. We considered phonemes and their durations as representative features of speech rhythm and used them as input for the speaker identification model. The performance of the proposed method was moderate in speaker identification, though it improved when the number of speakers included in the database increased. We also verified that our proposed method can model phoneme duration accurately. The proposed method predicts phoneme duration more accurately than the conventional x-vector based on acoustic features and is comparable to a one-hot vector that requires retraining with a few utterances. The analysis of the embedding space showed that modeling speech rhythm enabled similar embeddings to be extracted from utterances with similar speaking rhythms. Future work includes the application of other DNN speech synthesis methods such as the duration predictor in FastSpeech2~\cite{ren2021fastspeech} and AlignTTS~\cite{zeng2020aligntts}, as well as higher quality end-to-end speech synthesis (e.g., Tacotron2~\cite{shen2018natural}) in which both the proposed and x-vectors are input as auxiliary information.

%\section*{Acknowledgments}

\bibliographystyle{ieicetr}
\bibliography{myrefs}

\end{document}